# Development of a turbulent spot into a stripe pattern in plane Poiseuille flow

**Hiroshi Aida, Takahiro Tsukahara\*, and Yasuo Kawaguchi**
Department of Mechanical Engineering, Tokyo University of Science
Yamazaki 2641, Noda-shi, Chiba, 278-8510 Japan
\*E-mail: tsuka@rs.tus.ac.jp

**ABSTRACT**

A structure consisting of quasi-laminar and turbulent regions in a stripe pattern, which can be found in a transitional plane channel flow, is called 'turbulent stripe'. In the previous works, the emergence of this structure was confirmed only in the case of decreasing Reynolds number from turbulent regime. In the present study, its formation from a turbulent spot has been investigated using a direct numerical simulation in a relatively large-scale computational domain of $L_x \times L_y \times L_z$ = 731.4$\delta$ × 2$\delta$ × 365.7$\delta$ at $Re_\tau$ = 56 and $L_x \times L_y \times L_z$ = 640$\delta$ × 2$\delta$ × 320$\delta$ at $Re_\tau$ = 64. We observed the stripe pattern of quasi-laminar and turbulent regions inside of the spot. However, the turbulent eddies decayed more rapidly for $Re\tau$= 56 than those for $Re_\tau$ = 64. The developed spot at $Re_\tau$ = 64 was found to be different form from that at $Re_\tau$ = 56.

**INTRODUCTION**

Through direct numerical simulation (DNS), Tsukahara *et al.* (2005, 2007, 2009) found a structure consisting of quasi-laminar and turbulent regions in the stripe arrangement in transitional plane channel flows with $Re_\tau = u_\tau\delta/\nu \leq 80$ ($Re_\tau$ is the friction Reynolds number, $u_\tau$ is the friction velocity, $\delta$ is the channel half width, and ν is the kinematic viscosity). This structure was named turbulent stripe. The emergence of this structure was also verified experimentally by Hashimoto *et al.* (2009) for the bulk Reynolds number ranging from $Re_m = 2u_m\delta/\nu$= 1700 to 2000, where $u_m$ is the bulk mean velocity. Such localized, disordered motion is similar to a turbulent equilibrium puff in a transitional pipe flow. While a puff has been referred to as an incomplete relaminarization process (Wygnanski *et al.*, 1973; 1975), a slug in a pipe was associated with the transition from laminar to turbulent flow. As for the channel flow, the emergence of the turbulent stripe has been confirmed in previous studies only in the case of decreasing Reynolds number (using a fully-developed turbulent flow at a moderate Reynolds number as an initial condition). Recently, Duguet *et al.* (2010) observed the flow-field development into the turbulent stripe in a plane Couette flow that was triggered by initial random disturbances. Their DNS studies further demonstrated that a spreading spot in the transitional regime reproduces spatio-temporal intermittency structures, such as turbulent bands. This work points out that turbulent stripe always emerge in the transitional plane Couette flow without hysteresis.

In the plane Poiseuille flow, it can be expected that the turbulent stripe should develop from a turbulent spot in the transitional Reynolds-number regime. Nonetheless, there have been few attempts to track the growth of a single turbulent spot over a long time. Therefore, we performed DNSs of growth of a turbulent spot in the plane Poiseuille flow. In this paper, although the present work is a continuation of our previous DNS (Aida *et al.*, 2010), we report additional DNS results and discuss the formation of turbulence pattern from a spot with emphasis on its Reynolds-number dependence.

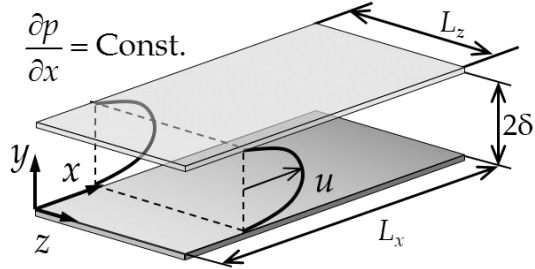

Figure 1. Configuration of channel flow.

**NUMERICAL PROCEDURE**

The mean flow under consideration was a plane Poiseuille flow driven by a uniform pressure gradient, as shown in Fig. 1. The periodic boundary condition was imposed in the horizontal (*x* and *z*) directions and the non-slip condition was applied on the walls. The coordinates and flow variables were normalized by $u_\tau$, $\delta$, and $\nu$: in this paper, quantities with the superscript of $*$ indicate those normalized by $\delta$, and the superscript of $+$ indicates the wall-unit normalization. The fundamental equations were the continuity equation and the Navier-Stokes equation.

For the spatial discretization, the finite difference method was adopted. The numerical scheme with the 4th-order accuracy was employed in the streamwise and spanwise directions, while the one with the 2nd-order was applied in the wall-normal (*y*-) direction. Time advancement was executed by a semi-implicit scheme: the 2nd-order Crank-Nicolson method for the viscous term in the wall-normal direction and the 2nd-order Adams-Bashforth method for the other terms. Uniform grid mesh was used in the horizontal directions, and non-uniform mesh in the wall-normal direction.

The friction Reynolds number was fixed to be 56 and 64, at which the turbulent stripe was found to emerge spontaneously in an initially fully turbulent flow (when the Reynolds number was decreased from the turbulent regime). Other detailed numerical conditions are summarized in Table 1. We employed relatively large-scale computational domains of $L_x \times L_y \times L_z = 731.4\delta \times 2\delta \times 365.7\delta$ at $\mathrm{Re}_\tau = u_\tau\delta/\nu = 56$ and of $L_x \times L_y \times L_z = 640\delta \times 2\delta \times 320\delta$ at $\mathrm{Re}_\tau = 64$ with $4096 \times 64 \times 2048$ grids. Such a large size of the domain allows us to consider a growing spot is not affected by neighboring spots resulting from the periodic boundary conditions, at least in the early stage of its spatial development.

Table 1. Computational conditions: $L_i$, domain size; $N_i$, number of grids; $\Delta x_i$, grid resolution; and $\Delta t$, time step.

| $Re_\tau$ | $L_x \times L_y \times L_z$ | $N_x \times N_y \times N_z$ | $\Delta x^+, \Delta z^+$ | $\Delta y^+_{max}$ | $\Delta y^+_{min}$ | $\Delta t^+$ |
|---|---|---|---|---|---|---|
| 56 | $731.4\delta \times 2\delta \times 365.7\delta$ | $4096 \times 64 \times 2048$ | 10.0 | 3.54 | 0.295 | 0.0224 |
| 64 | $640\delta \times 2\delta \times 320\delta$ | $4096 \times 64 \times 2048$ | 10.0 | 4.05 | 0.337 | 0.0128 |

A laminar flow field was used as the initial condition. The turbulent spot was triggered by a vortex pair, which had the following analytical form:

$$\begin{cases} \Psi = A(1-y^2)^2 z \cdot \exp(-x^2 - z^2) \\ u = 0 \\ v = \Psi_z \\ w = -\Psi_y \end{cases} \quad (1)$$

where *A* is an amplitude coefficient that provides the maximum initial wall-normal velocity comparable to the magnitude of center-line streamwise velocity of the laminar flow. This simple double-vortex disturbance was employed by Henningson & Kim (1991). It has been known from experimental investigations of spots in various types of flows that the spot characteristics become essentially independent of the initial disturbance if its magnitude is strong enough to develop.

**RESULT AND DISCUSSION**

*Instantaneous Velocity*

Figure 2 shows instantaneous distributions of wall-normal velocity ($v^+ = v/u_\tau$) in the ($x^*$, $z^*$)-plane at the channel center, presenting a temporal evolution of a turbulent spot. Figure 2(a) shows the flow field at $Re_\tau$ = 56. We observed that, in the first stage of development, the vortex pair as the initial disturbance broke down and developed into a well-known arrowhead-shaped turbulent spot as it propagated downstream (see Fig. 2(a)-1). In the downstream of the spot, a disturbed but non-turbulent region can be found, and the turbulent eddies were preceded by oblique waves. These findings are in consistent with those observed by Carlson *et al.* (1982) and Henningson *et al.* (1987). The spot changed its form into a V-shape at $t^+= tu_\tau/\nu = 800$ (Fig. 2(a)-2). At $t^+= 1600$, quasi-laminar regions emerged in the V-shaped turbulent region. Quasi-laminar and turbulent regions formed multiple V-shapes. After that, we can see the stripe arrangement of quasi-laminar and turbulent regions like a turbulent stripe inside of the spot (Fig. 2(a)-3). Turbulent regions began to branch at the edge of the spot, and the branching regions grew parallel to each other obliquely to the streamwise direction at an angle of 20–25 degrees (Fig. 2(a)-4). Figure 2(b) also shows a temporal evolution of a spot but for $Re_\tau$= 64. The spot at $t^+= 400$ also changed arrowhead-shape like the spot at $Re_\tau$ = 56 (Fig. 2(b)-1). However, the turbulent eddies hardly ever decayed in the downstream region at $t^+$ = 800 (Fig. 2(b)-2). Moreover, the turbulent region developed significantly in the upstream region as well as downstream region and changed into a different form from the spot at $Re_\tau$ = 56. At $t^* = 3200$, the intermittent turbulent region existed inside the spot but few branching turbulent regions can be seen at the edge of the spot. It should be noted that, at $Re_\tau$= 56, turbulence decayed and quasi-laminar regions emerged more significantly compared with the case at $Re_\tau$= 64. It allowed the spot to change into several bands of turbulence and each turbulent bands developed parallel to each other with preceded by oblique waves.

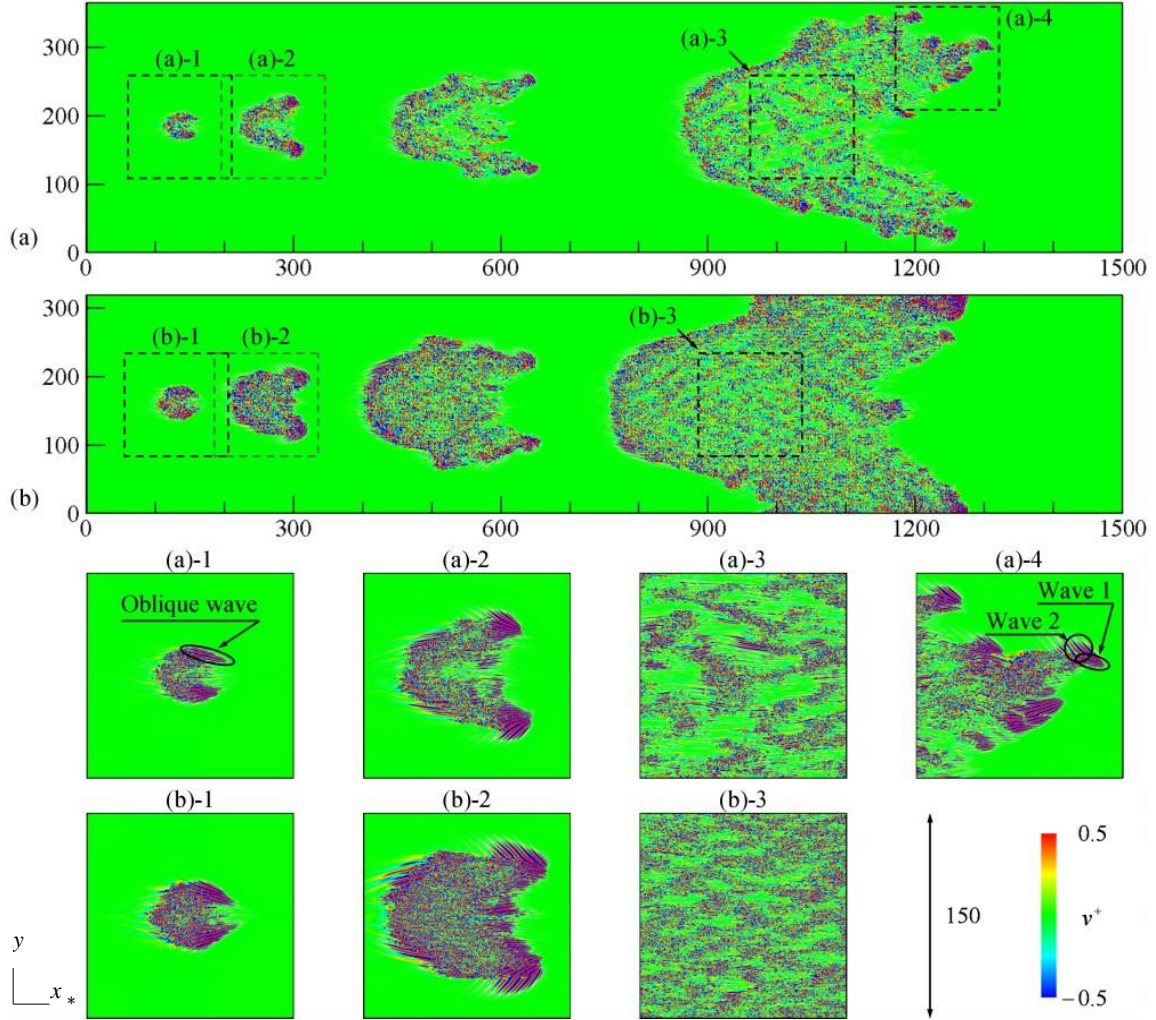

Figure 2. Contours of instantaneous wall-normal velocity in the $(x^*, z^*)$-plane at the channel center at $t^+$= 400, 800, 1600 and 3200: (a) $Re_\tau$ = 56, (b) 64. From (a)-1 to (a)-4 and from (b)-1 to (b)-3 are the enlarged views of the region within the dashed box in (a) and (b), respectively.

*Spreading of Disturbance*

We defined three points on the spot to discuss the propagation velocity of the spot. Point $x_d^*$ represents the streamwise position of the front (downstream) interface of the spot, point $x_u^*$ the rear (upstream) interface, and point $z^*$ the spanwise interfaces (refer to Fig. 3). The propagation of the spot was plotted as a function of time ($t^+$) in Fig. 4. As shown in Fig. 4, all points are seen to fall on straight lines, showing that their respective spot features propagate at constant speeds although turbulent region splits into quasi-laminar and turbulent regions inside of the spot. The propagation velocity of $x_u^*$ at $Re_\tau$ = 64 is slower than that at $Re_\tau$ = 56, although the propagation velocities of $x_d^*$ and $z^*$ are almost independent of the Reynolds number. It means that the spot at $Re_\tau$ = 64 developed more significantly in the upstream region than at $Re_\tau$ = 56, and it results in the different form from the V-shape spot at $Re_\tau$ = 56.

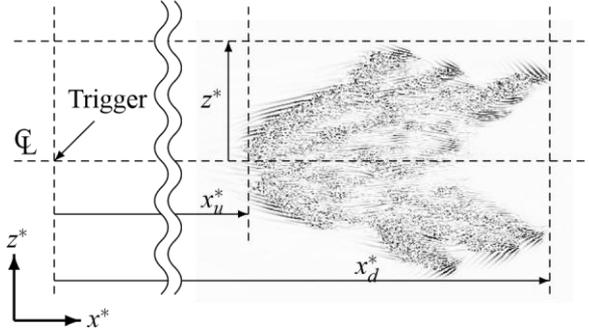

Figure 3. Turbulent-spot nomenclature.

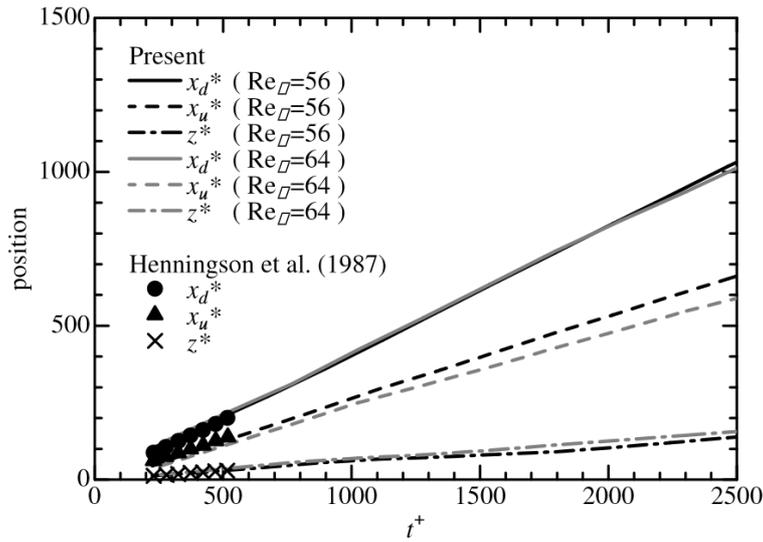

Figure 4. Position of the spot features: $x_d^*$ and $x_u^*$, front and rear interfaces, respectively; z, spanwise interface.

*Time Evolution of Turbulent Regions*

Toh & Itano (2005) traced the spanwise movement and generation of both near-wall and large-scale structures by plotting time evolution of the spanwise locations of low-speed regions. They demonstrated that, when an interval between near-wall streak structures exceeds some critical length of the order of 100 wall units, a new wall streak emerges between the separated streaks. In our study, a similar behavior can be found in large-scale motions of the order of about $20\delta$, that is, a new turbulent region emerged spontaneously from quasi-laminar region between the separated turbulent bands as the spot developed. This is illustrated in Fig. 5. The contour shows the spanwise distribution of the wall-normal velocity at the center of the spot ($x^* = (x_d^* + x_u^*)/2$, refer to Fig. 3) and the channel center ($y^* = 1$), with time in the vertical axis. At $t^+= 0$, a turbulent region emerged with an initial disturbance. Growing to a size of 20δ in the spanwise direction, the localized turbulent region split into two parts and a quasi-laminar region emerged between them in Fig. 5(a). The quasi-laminar region spread and then a new localized turbulent region emerged inside the quasi-laminar region. It can be said that when one (either laminar or turbulent) region develops larger in the spanwise direction than some threshold, the other region emerges to keep the spanwise spacing of each region almost constant. At $Re_\tau = 64$ (Fig. 5(b)), the quasilaminar and turbulent regions also emerged as similar to those at $Re_\tau = 56$. However, the turbulent regions for the higher Reynolds number spread more actively in the spanwise direction. When compared the spot shapes between the different Reynolds numbers (see Figs. 2(a) and 2(b)), its outline for $Re_\tau = 64$ seems to be circular, or crescent shape as described by Henningson & Kim (1991), rather than the V-shape. Moreover, the size of the turbulent region at the downstream edge of the spot was larger compared to that at $Re_\tau = 56$.

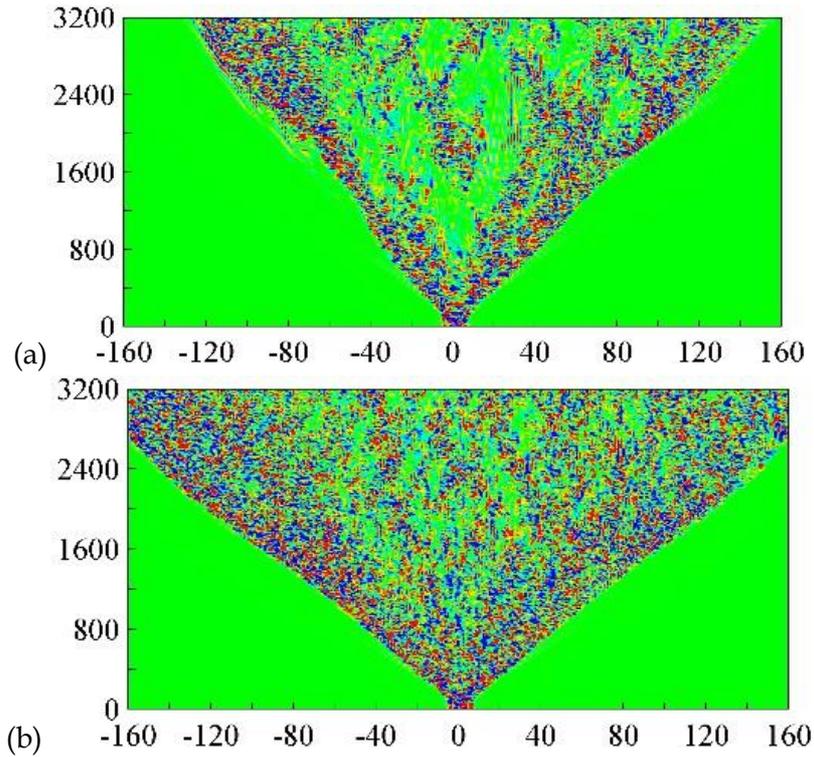

Figure 5. Space-time ($z^*$-$t^+$) plot of wall-normal velocity ($v^+$) showing the formation of a turbulent-laminar pattern: (a) $Re_\tau = 56$; (b) $Re_\tau = 64$.

*Vortex Structure*

Figure 6 shows the second invariant of deformation tensor:

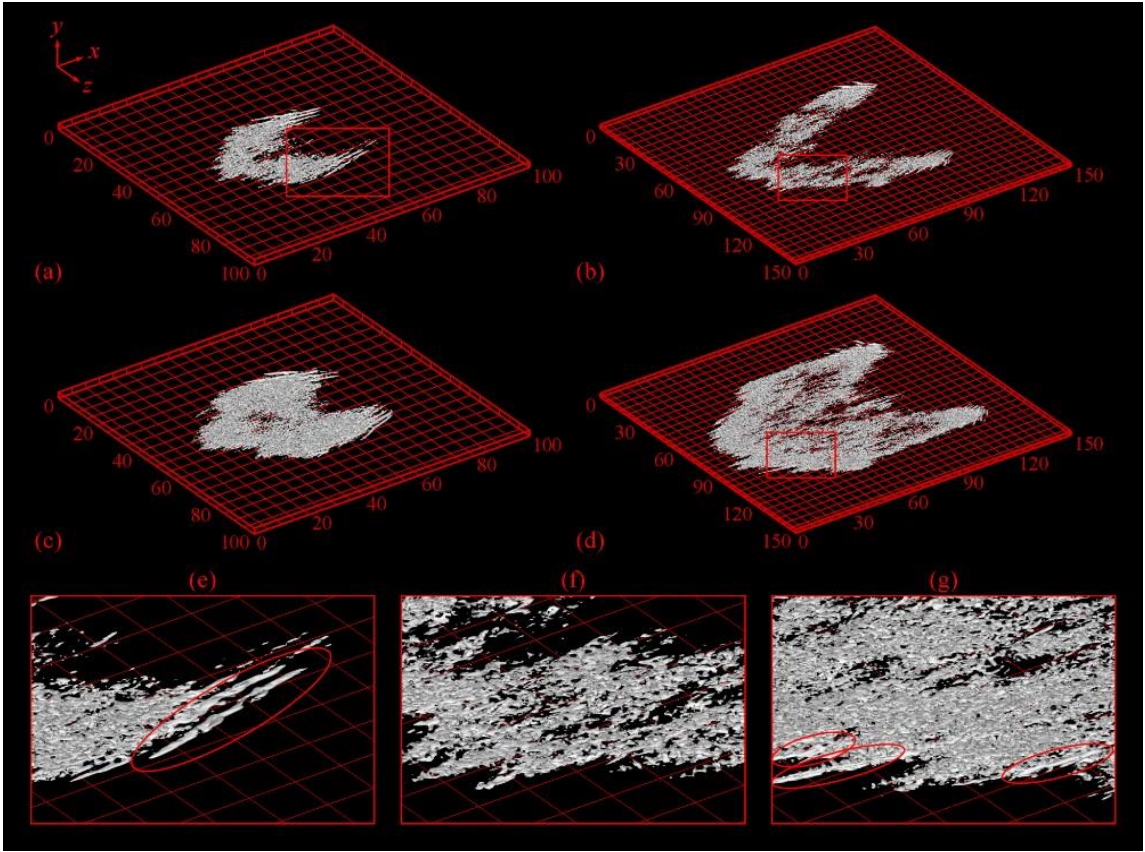

Figure 6. Iso-surfaces of second invariant of deformation tensor (*II*), which is equivalent to the vortical position. The direction of the mean flow is from bottom-left to top-right. (a) $t^+$ = 400, (b) 800 at $Re_\tau$ = 56; (c) $t^+$ = 400, (d) 800 at $Re_\tau$ = 64. (e) and (f) are the enlarged views of the region within the box in (a) and (c).

$$\text{II} = \frac{\partial u_i^+}{\partial x_j^*}\frac{\partial u_j^+}{\partial x_i^*}. \qquad (2)$$

It can be easily found that no vortex can be seen in the downstream of the spot and vortex cluster takes a V-shape. Figure 6(e)–(g) are the enlarged views of the regions within the box in (a), (b) and (d). In Fig. 6(e), elongated vortices at the wing-tip regions were found, and they were staggered in (*y*-*z*)-plane. It is considered that the staggered alignment of the vortices induced the oblique waves and make their wavelength to be about twice of the channel height as reported by Alavyoon *et al.* (1986). There could also be seen a few elongated vortices in the upstream region of the spot for $Re_\tau$ = 64 (Fig. 6(g)) but they are absent for $Re_\tau$ = 56 (Fig. 6(f)). For both cases, the oblique waves obviously appeared at the downstream of the spot, as seen in Fig. 6(e). However, the structure of the upstream oblique waves seem different from the downstream one. Sparse regions can be found inside the V-shape spot, as shown in Fig. 6(b) and (d). These regions developed into quasi-laminar regions inside the spot.

*T. Tsukahara et al.*

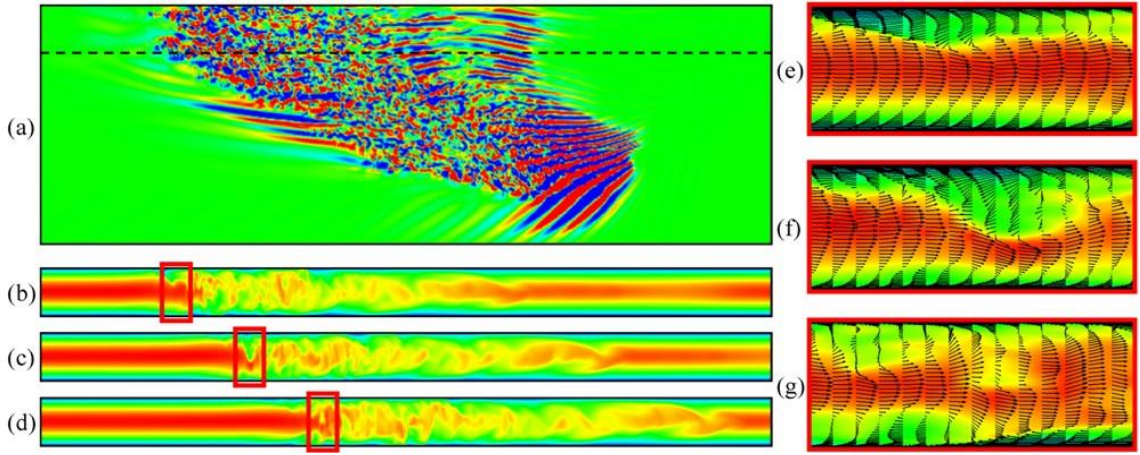

Figure 7. (a) Same as Fig. 2(a)-2, but for the visualized volume is 150δ × 50δ. (b)–(d) Contours of instantaneous streamwise velocity in the ($x^*$, $y^*$)-plane at the positions of the dashed line in (a), at $t^+$ = 800, 850, and 900. Contour color ranges from blue to red [0, 28]. The visualized volume is 150δ × 2δ, but aspect ratio of $x$ to $y$-direction is 0.2. (e)-(g) are the enlarged views of the region within the red box in (b)-(d) with vectors showing streamwise ($u^+ - u_m^+|_{laminar}$) and wall-normal ($v^+$) velocities, but aspect ratio of $x$ to $y$-direction is 1.

*Mechanism of Transition*

Figures 7(b)–(d) show temporal changing of the streamwise velocity in the ($x^*$, $y^*$)-plane at different instants. High-speed fluid (of the surrounding laminar flow) impinges on downstream low-speed fluid (of the turbulent spot) and curved into one side of the channel, as shown in (b). Low-speed flow near the other side of the walls was induced into the channel center: see the red box in (c). Figures 7(e)–(g) are the enlarged views of the region within the red box in (b)-(d), respectively, with vectors showing streamwise ($u^+ - u_m^+|_{laminar}$) and wall-normal ($v^+$) velocities. Here, $u_m^+|_{laminar}$ denotes the bulk mean velocity of laminar flow. The velocity gradient in the channel-central region is increased by low-speed fluid blowing into the channel center. Then due to the Kelvin-Helmholtz instability, a series of vortices occurs, as shown in Fig. 7(d), and induces the transition to turbulence near the interface, but inside the spot. This aspect is similar to the driving mechanism of a turbulent puff (Shimizu & Kida, 2009).

At another position of the upstream spot interface that is accompanied by the oblique waves, an oncoming high-speed flow region is curved and meandered in the $y$ direction more significantly (figure not shown), compared with that observed in Fig. 7. It is conjectured that the high-speed motion approaching the spot interface should be decelerated with giving rise to the oblique waves. Here, the existence of oblique waves can be interpreted as the appearance of longitudinal vortex packets. As shown in Fig. 6(g), a few of elongated streamwise vortices are, indeed, clearly confirmed in the upstream interface of the spot for $Re_\tau$ = 64, while they are absent or very weak for $Re_\tau$ = 56. It implies that the momentum transfer from the mean to these vortices is large in the higher Reynolds-number flow. Therefore, the transition at the upstream region of the spot at $Re_\tau = 56$ is slower than that at $Re_\tau = 64$.

**CONCLUSION**

In the present study, we performed large-scale DNS on a turbulent spot developing into turbulent stripes in a channel flow in the transitional Reynolds-number regime (at the friction Reynolds numbers of $Re_\tau = 56$ and 64).

We found that the spot developed in the form of stripe arrangements of quasi-laminar and turbulent regions inside of the spot, while keeping each region almost constant in scale. We observed different oblique waves at the edge of the developed spot from those observed in the arrowhead-shaped one. The turbulent regions also exist intermittently inside a spot at $Re_\tau = 64$. However, the Reynolds-number dependences of the development- and decay-rates of turbulent region, especially the highly disturbed region at the upstream interface of the growing spot, allows the spot itself to be different forms.

**ACKNOWLEDGMENT**

The present computations were performed with the use of the supercomputing resources at Cyber-Science Center of Tohoku University. This work has been supported by KAKENHI (#22760136).

This paper is a revised and expanded version of a paper entitled "Development process of a turbulent spot into stripe pattern in plane Poiseuille flow," presented by H. Aida, T. Tsukahara, and Y. Kawaguchi, at the Seventh International Symposium on Turbulence and Shear Flow Phenomena, Ottawa, Canada, 28-31 July 2011, Paper-ID #8D2P (USB), 6 pages.